\documentclass[12pt]{article}
\usepackage[final]{graphics}
\usepackage{amsfonts,amsbsy}
\def\empile#1\over#2{\mathrel{\mathop{\kern 0pt#1}\limits_{#2}}}

\catcode`\@=11

\newcount\@tempcntc
\def\@citex[#1]#2{\if@filesw\immediate\write\@auxout{\string\citation{#2}}\fi
  \@tempcnta\z@\@tempcntb\m@ne\def\@citea{}\@cite{%
        \@for\@citeb:=#2\do%
    {\@ifundefined{b@\@citeb}%
        {\@citeo\@tempcntb\m@ne\@citea%
                \def\@citea{,\penalty\@m\ }{\bf ?}\@warning%
                {Citation `\@citeb' on page \thepage \space undefined}}%
        {\setbox\z@\hbox{\global\@tempcntc0\csname b@\@citeb\endcsname\relax}
     \ifnum\@tempcntc=\z@ \@citeo\@tempcntb\m@ne%
       \@citea\def\@citea{,\penalty\@m}%
       \hbox{\csname b@\@citeb\endcsname}%
     \else%
      \advance\@tempcntb\@ne%
      \ifnum\@tempcntb=\@tempcntc%
      \else\advance\@tempcntb\m@ne\@citeo%
      \@tempcnta\@tempcntc\@tempcntb\@tempcntc\fi\fi}}\@citeo}{#1}}%

\def\@citeo{\ifnum\@tempcnta>\@tempcntb\else\@citea
  \def\@citea{,\penalty\@m}%
  \ifnum\@tempcnta=\@tempcntb\the\@tempcnta\else
   {\advance\@tempcnta\@ne\ifnum\@tempcnta=\@tempcntb \else
\def\@citea{--}\fi
    \advance\@tempcnta\m@ne\the\@tempcnta\@citea\the\@tempcntb}\fi\fi}

\catcode`\@=12

\global\mathchardef\lesssim "142E

\newcommand{\slL}{\raise.15ex\hbox{$/$}\kern-.53em\hbox{$L$}}
\newcommand{\slP}{\raise.15ex\hbox{$/$}\kern-.67em\hbox{$P$}}
\newcommand{\slp}{\raise.1ex\hbox{$/$}\kern-.63em\hbox{$p$}}
\newcommand{\slq}{\raise.1ex\hbox{$/$}\kern-.63em\hbox{$q$}}
\newcommand{\slv}{\raise.1ex\hbox{$/$}\kern-.63em\hbox{$v$}}
\newcommand{\slR}{\raise.15ex\hbox{$/$}\kern-.53em\hbox{$R$}}
\newcommand{\slQ}{\raise.15ex\hbox{$/$}\kern-.53em\hbox{$Q$}}
\newcommand{\slK}{\raise.15ex\hbox{$/$}\kern-.53em\hbox{$K$}}
\newcommand{\slk}{\raise.15ex\hbox{$/$}\kern-.53em\hbox{$k$}}
\newcommand{\slSigma}{\raise.15ex\hbox{$/$}\kern-.53em\hbox{$\Sigma$}}
\newcommand{\slcalP}{\raise.15ex\hbox{$/$}\kern-.63em\hbox{$\cal P$}}
\newcommand{\slA}{\raise.15ex\hbox{$/$}\kern-.73em\hbox{$A$}}
\newcommand{\slbfA}{\raise.15ex\hbox{$/$}\kern-.73em\hbox{${\boldsymbol A}$}}
\newcommand{\slpartial}{\raise.15ex\hbox{$/$}\kern-.53em\hbox{$\partial$}}
\def\p{{\boldsymbol p}}
\def\l{{\boldsymbol l}}
\def\f{{\boldsymbol f}}
\def\b{{\boldsymbol b}}

\newcommand\Eq[1]{Eq.~(\ref{#1})}

\def\gs{g_{\rm s}}

\def\Nf{N_{_{\rm F}}}
\def\Cf{C_{_{\hspace{-0.1em} \rm F \hspace{0.1em}}}}
\def\Nc{N_{\rm c}}
\def\mD{m_{_{\rm D}}}
\def\gammaE{\gamma_{_{\rm E}}}
\def\Meff{M_{\rm eff}}
\def\alphas{\alpha_{\rm s}}
\def\alphaEM{\alpha_{_{\rm EM}}}
\def\Im{\, {\rm Im} \:}
\def\Re{\, {\rm Re} \:}
\def\gsim{\mbox{~{\raisebox{0.4ex}{$>$}}\hspace{-1.1em}
        {\raisebox{-0.6ex}{$\sim$}}~}}

\def\PiR{\Pi_{_{\rm R}}}
\def\PiL{\Pi_{_{\rm L}}}
\def\seq{\!\! = \!\!}

\newcommand{\be}{\begin{equation}}
\newcommand{\ee}{\end{equation}}
\newcommand{\bea}{\begin{eqnarray}}
\newcommand{\ena}{\end{eqnarray}}

\def\build#1\over#2{\mathrel{\mathop{\kern 0pt#1}\limits_{#2}}}

\newcount\toto
\def\strip[hep-ph/0204#1]{#1}
\def\addoneto#1{\toto=#1\relax%
        \global\advance\toto by 1\relax%
        \the\toto}
\def\subtractoneto#1{\toto=#1\relax%
        \global\advance\toto by -1\relax%
        \the\toto}

\font\tenimbf=cmmib10 at 10pt
\font\sevenimbf=cmmib10 at 7pt
\font\fiveimbf=cmmib10 at 5pt
\newfam\imbf
\textfont\imbf=\tenimbf
\scriptfont\imbf=\sevenimbf
\scriptscriptfont\imbf=\fiveimbf

\begin{document}
\title{\bf{Landau-Pomeranchuk-Migdal resummation for dilepton
production\footnote{A test program calculating the LPM corrections to
the photon and dilepton rates can be found at the URL: {\tt
http://www-spht.cea.fr/articles/T02/150/libLPM/}}}}
\author{P.~Aurenche$^{(1)}$, F.~Gelis$^{(2)}$, G.D.~Moore$^{(3)}$,
H.~Zaraket$^{(4)}$} \maketitle
\begin{center}
\begin{enumerate}
\item Laboratoire d'Annecy-le-Vieux de Physique Th\'eorique, B.P. 110,\\
UMR 5108 du CNRS associ\'ee \`a l'Universit\'e de Savoie,\\
74941 Annecy-le-Vieux Cedex, France
\item Service de Physique Th\'eorique, Bat. 774, CEA/Saclay,\\
91191, Gif-sur-Yvette Cedex, France
\item Department of Physics, University of Washington,\\
Seattle, Washington 98195, USA\\
{\sl and}\\
Department of Physics, McGill University,\\
3600 rue University, Montr\'{e}al QC H3A 2T8, Canada
\item Physics Department, University of Winnipeg,\\
Winnipeg, Manitoba R3B 2E9, Canada
\end{enumerate}
\end{center}

\begin{abstract}
We consider the thermal emission rate of dileptons from a QCD plasma in
the small invariant mass ($Q^2 \sim \gs^2 T^2$) but large energy 
($q^0 \gsim T$) range.  We derive an integral equation which resums
multiple scatterings to include the LPM effect; it is valid at
leading order in the coupling.  Then we recast it as a differential
equation and show a simple algorithm for its solution.  We present
results for dilepton rates at phenomenologically interesting energies
and invariant masses.
\end{abstract}
\vskip 4mm
\centerline{\hfill SPhT-T02/150, LAPTH-946/02}

\section{Introduction}

In the last few years progress has been achieved in understanding the
production of electromagnetic probes in a quark-gluon plasma in
equilibrium. In the case of a hard photon (which is defined as having an
energy larger than the temperature T of the plasma) the usual Compton
and annihilation processes discussed long ago \cite{KapusLS1,BaierNNR1}
are not the only leading-order mechanisms. It turns out that bremsstrahlung
of a quark in a plasma gives a leading contribution
\cite{AurenGKP1,AurenGKP2} and, more importantly, a crossed process from
bremsstrahlung actually dominates at a large enough energy
\cite{AurenGKZ1,Mohan2,SteffT1,AurenGZ4}. This process, sometimes
referred to as off-shell annihilation or annihilation with scattering, is
of type 3-$body \rightarrow$ 2-$body$ : the photon is produced via
quark-antiquark annihilation where one of the incoming quarks is put
off-shell by scattering in the plasma. However, the
formation time of the photon is comparable to, or larger than, the
average time between soft scatterings of a quark in the plasma (inverse
of the damping rate). This implies that multiple scattering plays an
important role in these photon emission processes \cite{AurenGZ2}. 

A correct treatment of multiple scattering, even after performing Hard
Thermal Loop (HTL) resummation \cite{BraatenPisarski},
involves the resummation of ladder diagrams with
effective propagators for both fermions and gluons.  This resummation
leads to an integral equation, which has been solved to give
the photon emission rate to leading order in the strong coupling $\gs$
\cite{ArnolMY1,ArnolMY2,ArnolMY3}. The resummation of higher loop
diagrams implements the Landau-Pomeranchuk-Migdal
\cite{LandaP1,LandaP2,Migda1} effect which has been much discussed
recently in the context of jet energy loss in a hot medium
\cite{BaierDPS1,BaierDMPS2,BaierDMPS3,BaierDMS1,Zakha1,Zakha2,Zakha3}. 
However, there are differences
between the approach followed here for photon production and
that used for jet quenching: we work consistently in the HTL approach,
using the actual distribution of moving charges in the plasma and
dynamical screening of interactions, rather than a model of static,
Debye screened charges.  Note that no
magnetic mass needs to be introduced in order to screen transverse
gluons, since ultrasoft divergences have been shown to cancel thanks to
cancellations between diagrams of different topologies \cite{ArnolMY1}
(see also \cite{LebedS1,LebedS2,CarriK1,CarriKP1}).

Preliminary hydrodynamical studies
\cite{Sriva1,SrivaS1,HuoviRR1,Chaud1,ChaudK1}, which take into account
the above mechanisms of photon production, show that the thermal rate
(including the contribution from the hot hadronic phase) should
clearly emerge above prompt photon sources (produced in the early
stages of the collisions) in a range from 1 GeV $-$ 3 GeV to 5 GeV at
RHIC and LHC.

The above developments are quite timely in view of the recent WA98
measurements at SPS \cite{Aggara1} and of the upcoming results at RHIC
where the photon spectrum in gold-gold collisions is being measured
\cite{Averb1}. There is however an uncertainty concerning the possibility of
separating the spectrum of direct photons from that of photons which are
decay products of resonances, in particular $\pi^0$'s. In the energy
range of interest the $\pi^0$ signal may be almost two orders of
magnitude above the direct photon rate and the subtraction of the
$\pi^0,\ \eta,\ \cdots$ background from the data is a formidable
experimental challenge.

In the mean time it is interesting to consider an alternative channel which
carries the same dynamical information as the real photon production
rate, but which suffers a
different, less important, background. This is the case of the production of
low mass lepton pairs at large momentum. The invariant mass of the pair should
be high enough to be above the background of Dalitz pairs from the
pions, but small enough that the rate remains appreciable. Typically one
expects the 
range 150 Mev $< \sqrt{Q^2} <$ 500 MeV to be realistic for our purposes
($\sqrt{Q^2}$ is the invariant mass of the lepton pair). Higher mass ranges
(above the vector meson masses) should also be explored.

This rate has already been calculated for the Drell-Yan process
\cite{McLerT1}, for the $2\to 2$ processes \cite{AltheR1,ThomaT2}, and
for bremsstrahlung and off-shell annihilation in the approximation where
the quark undergoes a single scattering in the medium
\cite{AurenGZ3}. The analysis of multiple scattering is clearly called
for. Compared to the real photon case, the results of \cite{AurenGZ3}
show increased technical difficulties associated with the much more
complicated analytic structure of the diagrams: more unitarity cuts
contribute and the cancellation of divergences between the various cuts
is a subtle affair. In accordance with general theorems of perturbation
theory \cite{CatanW1}, the calculated rate turns out to be finite
everywhere except at the $q \bar q$ threshold ($Q^2 = 4 M_\infty^2$,
$M_\infty$ the quark thermal mass) where there remain integrable (square
root and logarithm) singularities. 
However, this threshold is right where the formation time is the
longest, and rescattering effects are expected to be largest.  Therefore
the two loop analysis \cite{AurenGZ3} is not reliable here.
%
%
%
%

The object of this paper is to compute the imaginary part of the
retarded current-current correlator, at momentum $Q$ which is hard ($q^0
\gsim T$) but near the light cone ($Q^2 \sim \gs^2 T^2$), at leading
order in $\gs$.  
This current-current correlator is related to the dilepton production
rate, per lepton species, via%
\footnote
    {%
    at leading order in $\alphaEM$ and neglecting the lepton mass
    }
\cite{Bella1}
\begin{equation}
\frac{dN_{_{\ell^+ \ell^-}}}{d^4x d^4 Q} 
        = \frac{\alphaEM}{12 \pi^4 Q^2 (e^{q^0/T}-1)}
        \Im \Pi_{_{R}}{}^\mu_\mu (Q) \, . 
\end{equation}
The calculation requires a resummation of diagrams analogous to that for
real photon production.
We first review the integral equation already obtained \cite{ArnolMY1}
in the case of a real photon ($Q^2=0$).  Then we extend it to off shell
photons.  Besides minor kinematic changes to the treatment of
\cite{ArnolMY1}, this also requires inclusion of longitudinally
polarized virtual photons.  It is then
shown how to obtain an ordinary differential equation by going to
impact parameter space, where the equation is solved
numerically.  The result is more complete, and in some ways simpler,
than the two loop perturbative one.  We find that
multiple scattering effects completely remove the $q\bar{q}$ threshold
behavior found in the perturbative treatment; the dilepton spectrum is
nonsingular and smooth.  Finally, we discuss some
phenomenological applications.

\section{LPM corrections for longitudinal photons}

In this section, we leave aside the production of photons by Compton
scattering ($g q \rightarrow \gamma q$) and annihilation ($ q \bar q
\rightarrow \gamma g$) which appear at one loop in the perturbative
expansion of the HTL effective theory \cite{AltheR1,ThomaT2} and we
consider only the processes of type $q i \rightarrow \gamma q i $ and $q
\bar q i\rightarrow \gamma i$ where $i$ denotes a quark, an antiquark or
a gluon. These appear at the two-loop level but because of a strong
collinear enhancement, powers of $T^2/M_\infty^2 \sim 1/\gs^2$ are
generated so that they contribute at the leading order in the strong
coupling. Higher order diagrams with a ladder topology involve the same
enhancement mechanism [each new rung in the ladder brings a factor of
${\cal O}(\gs^2 /\gs^2)$] and therefore they also contribute at leading
order. The summation of these dominant diagrams leads to an integral
equation which we now discuss~\cite{ArnolMY1,ArnolMY2,ArnolMY3}.

\subsection{Reminder: transverse modes}

\begin{figure}[tbp]
\centerline{\resizebox*{8cm}{!}{\includegraphics{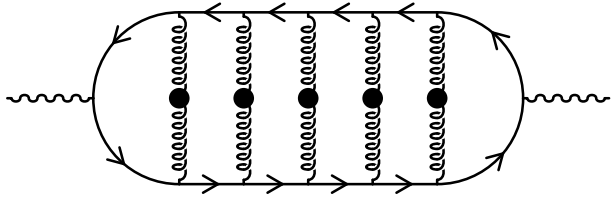}}}
\caption{\label{fig:ladder} Typical diagram which must be resummed to
determine $\Im \Pi_\mu^\mu$.  The gluon lines are soft and HTL-resummed,
the solid (quark) lines are hard, approximately on-shell, and resummed
to include the dominant imaginary part.}
\end{figure}

In \cite{ArnolMY1,ArnolMY2,ArnolMY3}, Arnold, Moore, and Yaffe summed
the set of diagrams of the form shown in Fig.~\ref{fig:ladder}, for lightlike
external momentum $Q$.  The lightlike condition allowed them to consider only
the two transverse polarizations; in general $\Pi^{\mu \nu}$ has
contributions from three polarizations (not four, because 
$Q_\mu \Pi^{\mu \nu}=0$), but when $Q^2=0$ the longitudinal polarization
also vanishes.  In terms of the retarded photon
polarization tensor, we can rewrite the result presented in these papers as
follows:
\begin{eqnarray}
\sum_{i=1,2}{\rm Im}\,\PiR{}_i^i(Q) & \approx & {{e^2
\Nc}\over{8\pi}}
\int_{-\infty}^{+\infty}dp_0\,
\;[n_{_{F}}(r_0)-n_{_{F}}(p_0)]\;
{{p_0^2+r_0^2}\over{(p_0r_0)^2}}
\nonumber\\
&&\qquad\;\;\qquad\times \Re \int {{d^2\p_\perp}\over{(2\pi)^2}}\;
2\p_\perp\cdot\f(\p_\perp)\; ,
\label{eq:AMY}
\end{eqnarray}
with $r_0\equiv p_0+q_0$, $n_{_{F}}(p_0)\equiv 1/(\exp(p_0/T)+1)$ the
Fermi-Dirac statistical weight, and where the dimensionless function
$\f(\p_\perp)$ obeys the following integral equation:
\begin{equation}
2\p_\perp=i\delta E \f(\p_\perp)+\gs^2 \Cf T \!\! \int
{{d^2\l_\perp}\over{(2\pi)^2}} \, {\cal C}(\l_\perp) \,
[\f(\p_\perp)-\f(\p_\perp+\l_\perp)]\; .
\label{eq:integ-f}
\end{equation}
In this integral equation, $\delta E\equiv q_0(\p
_\perp^2+M_\infty^2)/(2p_0r_0)$ is an energy denominator arising from
nearly on-shell quark propagators; it can be interpreted as the inverse
formation time of the photon.  Here $M_\infty^2\equiv \gs^2 \Cf
T^2/4$ is the thermal mass of a hard quark at leading order (or
equivalently, in the HTL approximation).
$\Cf = (\Nc^2-1)/2\Nc$ is the Casimir in the fundamental
representation of the gauge group $SU(\Nc)$; in QCD it is 4/3. 
The second term arises from
the gluon exchange lines in Fig.~\ref{fig:ladder} (both those shown
and those hidden in the fact that the quark line self-energies are
resummed).  The factor ${\cal C}(\l_\perp)$ is the kernel of a collision 
integral describing an elementary
collision of the emitter with another particle of the plasma.
In terms of
the transverse and longitudinal HTL self-energies $\Pi_{_{T,L}}(l_0/l)$
of the exchanged gluon, it is given by
\cite{ArnolMY1,ArnolMY2,ArnolMY3}:
\begin{eqnarray}
{\cal C}(\l_\perp) & \!\! \equiv \!\! &
\int
{{dl_0dl_z}\over{(2\pi)^2}} 2\pi\delta(l_0-l_z)\,{1\over{l_0}}\nonumber\\
&&\hspace{2.4em}\times
\sum_{\alpha=L,T}  {{2{\rm Im}\,\Pi_{\alpha}(L)}\over{(L^2-{\rm
Re}\,\Pi_{\alpha}(L))^2+({\rm Im}\,\Pi_{\alpha}(L))^2}}\;
P_{\alpha}^{\mu\nu}(L) \widehat{Q}_\mu \widehat{Q}_\nu\; ,
\label{eq:coll-term}
\end{eqnarray}
with $\widehat{Q}_\mu\equiv (1,{\boldsymbol q}/q)$ and $P_{_{T,L}}^{\mu\nu}(L)$
the transverse or longitudinal projector for a gluon of momentum $L$. In
\cite{AurenGZ4}, it was shown that this collision kernel can in fact
be calculated exactly thanks to a sum rule satisfied by the HTL gluon
propagator, leading to a very simple expression:
\begin{equation}
{\cal C}(\l_\perp)={1\over{\l_\perp^2}}-{1\over{\l_\perp^2+\mD^2}}\; ,
\end{equation}
where $\mD^2\equiv \gs^2 T^2 [\Nc+\Nf /2]/3$ is the Debye mass in
an $SU(\Nc)$ gauge theory with $\Nf $ fundamental representation flavors.

Generalizing the above equations to the case of the transverse modes of
a massive photon is straightforward, as one needs only to replace
$M_\infty^2$ in $\delta E$ by the following combination of the quark mass
and the photon invariant mass \cite{AurenGKP2,ArnolMY2}
\footnote{%
        This substitution is sufficient as long as we are considering
        hard photons that are not too virtual. It can be used if the
        following inequality is satisfied: 
        \begin{equation}
        Q^2\ll q_0^2\; .
        \end{equation}
        If this is not the case, then the kinematics is no longer
        dominated by small angle scatterings and all of the above
        equations must be reconsidered. Note that this inequality
        implies that one can interchange $q_0$ and the 3-momentum $q$
        at leading order, in all quantities where no cancellation
        ($q_0-q$) occurs.}:
\begin{equation}
    M_\infty^2\to \Meff^2\equiv M_\infty^2+
    {{Q^2}\over{q_0^2}}p_0 r_0\; .
\end{equation}
For $Q^2>4M_\infty^2$, this quantity can become negative and $\delta E$
can vanish. In this case, $\Meff^2$ should be understood as having
an infinitesimal imaginary part $i p_0 r_0 \epsilon / q_0$, as one can
readily see from its derivation from the quark energy denominators. 

This is not enough to fully calculate the dilepton production
rate; one must also include the longitudinal mode of the off shell
(virtual) photon.  We calculate the contribution of this mode in the next
subsection.

\subsection{Longitudinal mode}
The contribution of the longitudinal mode can be derived simply by
following a procedure similar to the one employed in
\cite{ArnolMY1,ArnolMY2}. Let us start by writing%
\footnote
    {%
    Note that, since the sum over polarizations
    $\sum_{\lambda}\epsilon_\lambda^\mu(Q)\epsilon_\lambda^{* \nu}(Q)$
    can be replaced by $-g^{\mu\nu}$ thanks to Ward identities, it is
    the {\sl opposite} of $\PiL$ that must be added to $\Pi_i{}^i$ in
    order to obtain the full $\Pi_\mu{}^\mu$.
    }:
\begin{equation}
\PiL(Q)\equiv\epsilon_{_{\rm L}}^\mu(Q) 
        \epsilon_{_{\rm L}}^{\nu *}(Q)\Pi_{\mu\nu}(Q)\; ,
\end{equation}
with the following longitudinal polarization vector
\begin{equation}
\epsilon_{_{\rm L}}^\mu(Q)\equiv{{(q,0,0,q_0)}\over{\sqrt{Q^2}}}\; .
\end{equation}
Making use of the Ward identity $Q^\mu\Pi_{\mu\nu}(Q)=0$ satisfied by
the photon polarization tensor, we obtain
\begin{equation}
\PiL(Q)={{Q^2}\over{q^2}}\Pi^{00}(Q)\; .
\label{eq:Pi_L}
\end{equation}
Therefore, we see that we need only the $00$ component of the photon
polarization tensor%
\footnote{
        To be consistent, if HTL fermion propagators are necessary
        then HTL corrections to the $q\bar{q}\gamma$ vertices are
        also needed \cite{AurenGZ3}.  However, the complete effect
        of this vertex is to ensure that the Ward identity is 
        satisfied, by rescaling $\Pi_{zz}$ by an $O(\gs^2)$ amount
        with respect 
        to $\Pi_{00}$.  By using the Ward identity to express 
        $\PiL$ in terms of $\Pi_{00}$, we have taken this into
        account, and can compute $\Pi_{00}$ at leading order only.}.
This also makes clear why $\PiL=0$ for real photons.

Now we must evaluate $\Pi^{00}(Q)$ summing over all diagrams of the type
shown in Fig.~\ref{fig:ladder}.
Recall that the retarded/advanced HTL propagator of a hard
quark can be approximated by:
\begin{equation}
S_{_{R,A}}(P)\approx i{{\overline{\slP}}\over{\overline{P}^2\pm
ip_0\epsilon}}\; ,
\end{equation} 
where $\overline{P}\equiv(p_0,\sqrt{\p^2+M_\infty^2}\hat{\p})$. 
It is easy to verify that one can replace the Dirac matrix
$\overline{\slP}$ in the numerator of the effective quark propagator by:
\begin{equation}
\overline{\slP}\to\epsilon(p_0)\; w(\epsilon(p_0) \, \p) \;
\overline{w}(\epsilon(p_0)\, \p)\; ,
\end{equation}
where we denote $w(\p) \!\equiv \! \sqrt{\omega_p/p}\; u(\p)$ with
$u(\p)$ the usual massless%
\footnote{%
        This property is true because
        the HTL resummation preserves the chiral invariance of the
        bare theory.} 
spinor normalized according to
$u(\p)\overline{u}(\p)=\slP$ 
(implicitly, $p_0>0$ in this relation)
and $\omega_p\equiv\sqrt{\p^2+M_\infty^2}$.  Then, one can associate the
spinors $w$ and $\overline{w}$ in the numerator of the propagator with
the vertices,
\begin{equation}
\gamma^\sigma\to
\underline{\gamma}^\sigma\: \equiv \: \overline{w}(\epsilon(p_0+l_0)(\p
+\l)) \: \gamma^\sigma \: w(\epsilon(p_0)\p)\; ,
\end{equation}
if $L\equiv(l_0,\l)$ is the momentum of the gauge boson entering
in the vertex.  For the vertices coupling the quark loop to the soft
gluons, we can simply neglect the gluon momentum, and write
$\underline{\gamma^\sigma}\approx 2P_\epsilon^\sigma$ with
$P_\epsilon\equiv(p_0,\epsilon(p_0)\p)$.  It is slightly more
complicated to deal with the $\gamma q\bar{q}$ vertices since the photon
can be hard, but we need only the $0$ component $\underline{\gamma}^0$, in
the configuration where the quark and the photon are nearly
collinear. In this limit, we find: $\underline{\gamma}^0\approx
2\sqrt{pr}\approx 2\sqrt{|p_0r_0|}$.

Now that all the spinor structure has been absorbed in the vertices,
performing the Dirac's trace is trivial; ${\rm
Tr}({\boldsymbol 1})=4$.  In fact, the vertices coupling the quark loop to the
gluons are the same as in a scalar theory, and the only differences
between scalar and fermionic particles lie in the coupling to the
photon and in the statistical weights. Therefore, we can
mimic%
\footnote{
        The substitution rules are the following: there is a $-$
        sign since it is the opposite of $\PiL$ that enters in
        $\Pi_\mu{}^\mu$. A factor $\epsilon(p_0r_0)$ comes from the last loop
        (other factors of $\epsilon(p_0r_0)$ are kept inside the function
        $g(\p_\perp)$). The factor $Q^2/q^2$ comes from the relationship
        between $\PiL$ and $\Pi_{00}$, and the factor
        $(p_0^2+r_0^2)/(p_0r_0)$ must be removed from Eq.~(\ref{eq:AMY}) since
        it was there to account for the coupling of the quark to the
        transverse modes of the photon.}  
Eq.~(\ref{eq:AMY}) for the retarded self-energy of
the longitudinal photon in order to obtain
\begin{eqnarray}
\Im \PiL(Q) & = & -{{e^2\Nc}\over{4\pi}}{{Q^2}\over{q^2}}
        \int\limits_{-\infty}^{+\infty}{dp_0}\,\;
        [n_{_{F}}(r_0)-n_{_{F}}(p_0)]
        {{\epsilon(p_0r_0)}\over{p_0r_0}}\nonumber\\
        && \hspace{6em} \times
\Re \int {{d^2\p_\perp}\over{(2\pi)^2}}\;
2\sqrt{|p_0r_0|}\, g(\p_\perp)\; ,
\end{eqnarray}
where $g(\p_\perp)$ is a dimensionless scalar function which
describes the resummed coupling of the quark to the longitudinal photon,
and obeys the following integral equation, in complete
analogy\footnote{Just replace the ``tree-level'' vertex 
$2\p_\perp$ (i.e. the starting point of the iteration) by
$2\sqrt{|p_0r_0|}$ and the function $\f(\p_\perp)$ by
$g(\p_\perp)$.} with Eq.~(\ref{eq:integ-f}),
\begin{equation}
2\sqrt{|p_0r_0|}=i\delta E {g}(\p_\perp)+\gs^2 \Cf T\int
{{d^2\l_\perp}\over{(2\pi)^2}} {\cal C}(\l_\perp)
[{g}(\p_\perp)-{g}(\p_\perp+\l_\perp)]\; .
\label{eq:integ-g}
\end{equation}

Combining the result of \cite{ArnolMY1} and the above formulas for the
longitudinal mode, we can write the equation for the
complete polarization tensor of a massive photon:
\begin{eqnarray}\label{eq:LPM-complete}
{\rm Im}\,\PiR{}_\mu^\mu(Q) & \!\!\!\approx\!\!\! & {{e^2 \Nc}\over{2\pi}}
        \int_{-\infty}^{+\infty}dp_0\,
        \;[n_{_{F}}(r_0)-n_{_{F}}(p_0)]\; \times 
        \\&& 
        \times \Re \int\! {{d^2\p_\perp}\over{(2\pi)^2}}
        \left[
        {{p_0^2+r_0^2}\over{2(p_0r_0)^2}}\p_\perp\cdot\f(\p_\perp)
        +{{1}\over{\sqrt{|p_0r_0|}}}{{Q^2}\over{q^2}}\,g(\p_\perp)
        \right]\; , \nonumber 
\end{eqnarray}
where $\f(\p_\perp)$ and $g(\p_\perp)$ respectively
obey Eq.~(\ref{eq:integ-f}) and Eq.~(\ref{eq:integ-g}). Note that this
expression is valid for one flavor of the emitting quark, with
electrical charge $e$ (but $\Nf $ flavors of quarks are taken into
account for the scattering centers, as can be seen from the expression
of the Debye mass). For the emission by $\Nf $ families  of quarks
with electrical charges $e_s$, one must substitute:
\begin{equation}
e^2\to \sum_{f=1}^{\Nf }e_f^2\; .
\end{equation}

\subsection{Born term}
Before solving numerically the above integral equations, it is
interesting to show that they contain the contribution of the Born term,
corresponding to the direct $q\bar{q}$ annihilation into a virtual
photon. Indeed, if we solve the equations (\ref{eq:integ-f}) and
(\ref{eq:integ-g}) at order 0 in the collision term, we obtain
respectively
\begin{equation}
\p_\perp\cdot \f(\p_\perp)=
-4 \p_\perp^2 \: {{p_0 r_0}\over {q_0}} \:
{{i}\over{\p_\perp^2+\Meff^2+i{{p_0
r_0}\over{q_0}}\epsilon}}
\end{equation}
and
\begin{equation}
g(\p_\perp)=-4 \sqrt{|p_0 r_0|} \: {{p_0 r_0}\over {q_0}}\:
{{i}\over{\p_\perp^2+\Meff^2+i{{p_0
r_0}\over{q_0}}\epsilon}}\; .
\end{equation}
These quantities yield a nonzero real part, when integrated over $\p$,
provided that
$\Meff^2<0$.  This occurs when $Q^2 > 4M_\infty^2$,
i.e. above the threshold for the process $q\bar{q}\to \gamma^*$.
Keeping only the real part and substituting these values in
Eq.~(\ref{eq:LPM-complete}), we get
\begin{eqnarray}
\Im \PiR{}_\mu^\mu(Q) & \seq & {{e^2
\Nc}\over{2\pi}}
\int_{-\infty}^{+\infty}dp_0\,\theta(-\Meff^2)
\;[n_{_{F}}(r_0)-n_{_{F}}(p_0)]\;
\nonumber\\
&& \hspace{5em} \times \left[Q^2
+M_\infty^2\frac{p_0^2+r_0^2}{r_0p_0}\right]\; ,
\end{eqnarray}
which is identical to the expression obtained in \cite{AurenGZ3} for the
Born contribution to the virtual photon retarded self-energy. Note that
the term proportional to $M_\infty^2$ can be seen as an effect of taking
into account the correct HTL vertex correction in the calculation in
this Born term. In the present approach, we did not need to use these
vertices explicitly; rather, we relied on them to ensure the Ward
identity, used in the calculation of the longitudinal contribution
(see the derivation of Eq.~(\ref{eq:Pi_L})).

Therefore, the only leading order terms that are not included in
Eq.~(\ref{eq:LPM-complete}) are the $2\to 2$ processes considered in
\cite{AltheR1,ThomaT2}. This extra contribution should be added to our
results in order to obtain the complete leading order dilepton spectrum.

\section{Formulation in impact parameter space}

It is possible to approach the integral equations analytically either in
an expansion in the collision term (the zero and first orders are
known), or to leading logarithmic order in small $\delta E / {\cal C}$;
but each method is accurate only in a limited kinematic regime.
In general it is necessary to treat the integral equation numerically.
This has been done for on-shell photon production, in momentum space, by
using a variational approach \cite{ArnolMY2,Sastr1}.

However, a simpler method can be obtained after going to impact
parameter space by a Fourier transform.  Doing so is made practical by
the fact that the collision term is known in closed form and has a
rather simple Fourier transform. The method was outlined in
\cite{AurenGZ4} in the case of real photon production, and can be
extended to the case of dileptons.

Let us first introduce the Fourier transforms:
\begin{eqnarray}
\f(\p_\perp)&\!\!\! \equiv \!\!\!& \int d^2\b\, e^{-i\p_\perp\cdot \b}\, 
\f(\b)\; ,\nonumber\\
g(\p_\perp)&\!\!\! \equiv \!\!\!& \int d^2\b\, e^{-i\p_\perp\cdot \b}\, 
        g(\b)\; ,
\end{eqnarray}
where we use the same letter to denote both a function and its Fourier
transform, as the argument should make obvious which one enters in a
particular expression. The functions $\f(\b)$ and $g(\b)$ 
obey simple partial differential equations:
\begin{eqnarray}
i{{q_0}\over{2p_0r_0}}(\Meff^2-\Delta_\perp)\f(\b
)+\gs^2\Cf TD(\mD b)\f(\b) & \seq & -2i\nabla_\perp
\delta(\b)\; ,\nonumber\\
i{{q_0}\over{2p_0r_0}}(\Meff^2-\Delta_\perp){g}(\b
)+\gs^2\Cf TD(\mD b){g}(\b) & \seq & 2\sqrt{|p_0 r_0|}\delta(\b)\; ,
\label{eq:ODE-1}
\end{eqnarray}
where we denote
\begin{equation}
D(\mD b)\equiv {1\over{2\pi}}\left[
\gammaE +\ln\left({{\mD b}\over{2}}\right)
+K_0(\mD b)
\right]\; ,
\end{equation}
with $K_0$ the modified Bessel function of the second kind and 
$\gammaE$
the Euler constant, $\gammaE \approx 0.577$. As we shall see later, the
inhomogeneous terms proportional to $\delta(\b)$ in these
equations determine uniquely the normalization of the solution by
imposing the small $b$ behavior of the solution. The quantities that are
needed in Eq.~(\ref{eq:LPM-complete}) in order to compute the photon
polarization tensor are then given by
\begin{eqnarray}
{\rm Re}\;\int {{d^2\p_\perp}\over{(2\pi)^2}}
 \p_\perp \cdot \f(\p_\perp) &\!\! = \!\!& \lim_{b\to 0^+}{\rm Im}\;
\nabla_\perp\cdot\f(\b)\; ,\nonumber\\
{\rm Re}\int {{d^2\p_\perp}\over{(2\pi)^2}}\,
{g}(\p_\perp) &\!\!= \!\!& \lim_{b\to 0^+}{\rm Re}\;
g(\b)\; .
\label{eq:ODE-2}
\end{eqnarray}

We must now determine the boundary conditions to impose on the solutions
of Eqs.~(\ref{eq:ODE-1}) in order to determine them uniquely. First, the
functions $\f(\b)$ and $g(\b)$ must remain bounded
when $|\b|\to +\infty$.  The most general large $b$ behavior is
the sum of an exponentially growing and and exponentially shrinking
solution; only the shrinking solution is allowed, so the required
boundary condition is
\begin{eqnarray}
\lim_{b\to +\infty} \f(\b) & \!\!=\!\! & 0\; ,\nonumber\\
\lim_{b\to +\infty} g(\b) & \!\!=\!\! & 0\; .
\label{eq:boundary-1}
\end{eqnarray}
Another boundary condition can be derived at $b=0$, from the small
distance behavior of the differential equations satisfied by $\f
(\b)$ and $g(\b)$. It is easy to check that
$D(\mD b)$ vanishes like $b^2 \ln(b)$ when $b$ approaches zero, so that the
term proportional to $D(\mD b)$ becomes irrelevant in this limit.
In fact, the dominant small $b$ behavior is completely determined by the
Laplacian and the inhomogeneous term. One finds
easily:
\begin{eqnarray}
\f(\b) & \!\! \empile{\approx}\over{b\to 0^+} \!\!& {2\over \pi} {{p_0
r_0}\over{q_0}} {{\widehat\b}\over {b}}\; + O(\b) \; ,\nonumber\\
g(\b) & \!\! \empile{\approx}\over{b\to 0^+} \!\!& {{2i}\over \pi} {{p_0
r_0}\over{q_0}}\sqrt{|p_0 r_0|} \ln(\mD b)\; + O(b^0) \; .
\label{eq:boundary-2}
\end{eqnarray}
Note that even if these small $b$ behaviors are singular, they do not
affect Eqs.~(\ref{eq:ODE-2}) since the singular part of $\f(\b
)$ is purely real, and the singular part of $g(\b)$ is purely
imaginary. Since the differential equations (\ref{eq:ODE-1}) are second
order affine equations, a solution is uniquely determined by two
boundary conditions. Therefore, Eqs.~(\ref{eq:ODE-1}), (\ref{eq:ODE-2})
together with the boundary conditions (\ref{eq:boundary-1}) and
(\ref{eq:boundary-2}) provide a complete reformulation of the original
problem.

These differential equations can be simplified further by noting that
rotational invariance in the transverse plane implies that $g(\b)$
depends only on the modulus $b$, and that $\f(\b)$ can be
written as $\f(\b)\equiv \b h(b)$. Therefore, we
have
\begin{eqnarray}
\Delta_\perp \f(\b) & \seq & \b \left[
h^{\prime\prime}(b)+3{{h^\prime(b)}\over{b}}
\right]\; ,\nonumber\\
\Delta_\perp g(\b) & \seq & g^{\prime\prime}(b)+{{g^\prime(b)}\over{b}}\; .
\end{eqnarray}
In terms of the new function $h(b)$, the first of Eqs.~(\ref{eq:ODE-2})
becomes
\begin{equation}
{\rm Re}\;\int {{d^2\p_\perp}\over{(2\pi)^2}}
 \p_\perp \cdot \f(\p_\perp)= 2 \lim_{b\to 0^+}{\rm Im}\;
h(b)\; ,
\label{eq:h_want}
\end{equation}
and the desired differential equation is
\begin{equation}
i\frac{q_0}{2p_0 r_0}\left( \Meff^2 - \partial_b^2 - 
        3 \frac{\partial_b}{b} \right) h(b) + \gs^2 \Cf TD(\mD b) h(b)
        = 0 \, ,
\label{eq:f_version2}
\end{equation}
with boundary condition 
\begin{equation}
h(b) \empile{\approx}\over{b\to 0^+} \frac{2 p_0 r_0}{\pi q_0 b^2}
        + O(1) \, .
\label{eq:bound_v2}
\end{equation}

\section{Numerical resolution}

General nonlinear ordinary differential equations with mixed boundary
conditions 
are usually solved with the ``overshoot-undershoot'' method.  However,
because our differential equations are linear, they can be solved by a
single application of a quadratures ODE solver (specific algorithms can
be found in \cite{PressTVF1}).  We will describe the procedure for
$h(b)$; the procedure for $g(b)$ is completely analogous\footnote{The
procedure described in the present paper is slightly different from the
one outlined in \cite{AurenGZ4}. The latter has also been checked to
work, albeit more slowly since it requires two runs of the ODE solver
instead of one.}.

What we want is the imaginary part, at the origin, of the exponentially
decaying solution of \Eq{eq:f_version2}, normalized to satisfy the boundary
condition at zero, \Eq{eq:bound_v2}.  This can be achieved by finding
the decaying solution, with arbitrary normalization, and scaling it to
have the desired small $b$ behavior (including rotation by a complex
phase to make the $1/b^2$ piece purely real).  To find the pure decaying
solution, it is sufficient to begin at large $b$ with arbitrary
(nonzero) initial data for $h(b)$ and $h'(b)$ and to evolve the
differential equation inwards towards $b=0$.  The solution which grows
exponentially at large $b$ shrinks as we evolve towards the origin, the
normalizable solution grows.

To determine what value of $b$ is
sufficient to suppress the wrong solution by a suitably large factor, we
can determine the approximate large $b$ behavior of \Eq{eq:f_version2}
by approximating $D$ to be constant (at large $b$ it only varies
logarithmically, $D(\mD b) \simeq (\gammaE + \log(\mD b/2))$) and
dropping the $3 \partial_b/b$ term.  The result is,
\begin{eqnarray}
h(b\gg 1/\mD) & \sim & c_+ e^{\lambda_+ \mD b} +
        c_- e^{\lambda_- \mD b} \, , \nonumber \\
\lambda_\pm & = & \pm \sqrt{ \Meff^2 - 
        i \frac{2\gs^2 \Cf Tp_0 r_0}{q_0} D(\mD b)} \, .
\label{eq:lambda}
\end{eqnarray}
We should choose initial data at a value of $b$, large enough that the
ratio of exponentials $e^{\lambda_+ \mD b} / e^{\lambda_- \mD b}$ is
abundantly larger than the desired accuracy.  It is also easy to use the
above equations to ensure the initial data for $h$ and $h'$ are almost
those of the shrinking solution, so the coefficient $c_+$ is initialized
close to zero.  (This procedure is not exact because of the 
$3\partial_b/b$ term we dropped, and the $b$ dependence of $D$ which we
neglected.)  Starting with this initial data, the ODE,
\Eq{eq:f_version2}, is evolved towards $b=0$ by standard methods, and the
coefficients of the $b^{-2}$ and $b^0$ behavior are extracted.

For the extraction of the $b^{-2}$ and $b^0$ behavior, it is helpful to
determine the small $b$ behavior a little more precisely.  At small $b$,
$D(\mD b) \sim b^2 \ln(1/\mD b)$ is negligible; but the $\Meff^2$ term
may be large enough, only to be negligible at quite small $b$, where
numerical error in the $b^{-2}$ term can pollute the $b^0$ term.  Hence
it is useful to match to the behavior neglecting the $D(\mD b)$ term but
including all other terms;
\begin{equation}
h(b \ll 1/\mD) \sim c_1 \frac{\Meff K_1(\Meff b)}{b} + c_2
        \frac{I_1(\Meff b)}{\Meff b} \, ,
\end{equation}
with $I_1$ and $K_1$ the modified Bessel functions of the first and
second kind (for $\Meff^2 < 0$ one should replace $K,I$ with $Y,J$ and
$\Meff$ with $|\Meff|$).  The desired quantity, \Eq{eq:h_want}, is
just $(4 p_0 r_0/\pi q_0) \: {\rm Im}\: c_2/c_1$.  (By working in
terms of $c_2/c_1$, we avoid needing to rescale the solution to
the appropriate normalization.)  It is also possible to
include the small effect of $D(\mD b)$ perturbatively.

The treatment of $g(h)$ is completely analogous; the asymptotic large
$b$ analysis is the same.  The differential
equation differs only in that the coefficient of the $\partial_b/b$ term is
1 instead of 3, so the small $b$ behavior is
\begin{equation}
g(b \ll 1/\mD) \sim c_1 K_0(\Meff b) + c_2 I_0(\Meff b) \, ,
\end{equation}
and the desired quantity, \Eq{eq:ODE-2}, is
$(2 p_0 r_0 \sqrt{p_0 r_0}/\pi q_0) \: {\rm Im}\: c_2/c_1$.

\section{Phenomenology}

\begin{figure}[htbp]
\centerline{
\resizebox*{!}{8cm}{\rotatebox{-90}{\includegraphics{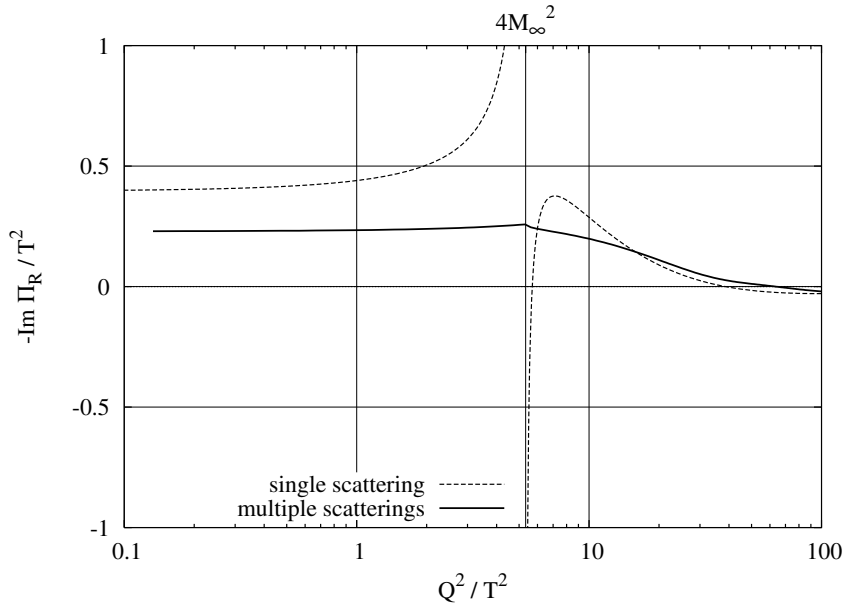}}}}
\caption{\label{fig:mass-spectrum} The sum of all the multiple 
  scattering diagrams compared to the single scattering contribution
  only. In this plot $\alpha_{\rm s}=0.3$ and $q_0/T=50$.}
\end{figure}

The major problem in the 2-loop evaluation of the dilepton rate was
the presence of a singularity at the location of the tree-level
threshold $Q^2=4M_\infty^2$. Basically, this divergence is a common
feature of any fixed loop order calculation near a phase-space
boundary. A first point to check in the present resummation is that
this problem has been cured. This is readily seen in the plot of
figure \ref{fig:mass-spectrum}.
In this plot, `single scattering' denotes the contribution calculated
in \cite{AurenGZ3}, i.e. what one would get by keeping only the first
order in an expansion of the integral of eqs.~(\ref{eq:integ-f}) and
(\ref{eq:integ-g}) in powers of the collision kernel ${\cal C}({\boldsymbol
  l}_\perp)$. One can see that including all orders in ${\cal C}({\boldsymbol
  l}_\perp)$ makes the result continuous at $Q^2=4M_\infty^2$. Note
that the Born term (i.e. the contribution of ${\cal C}^0$ in the
solution of the integral equations) is not included in this plot.

\begin{figure}[htbp]
\centerline{\resizebox*{!}{9cm}
        {\rotatebox{-90}{\includegraphics{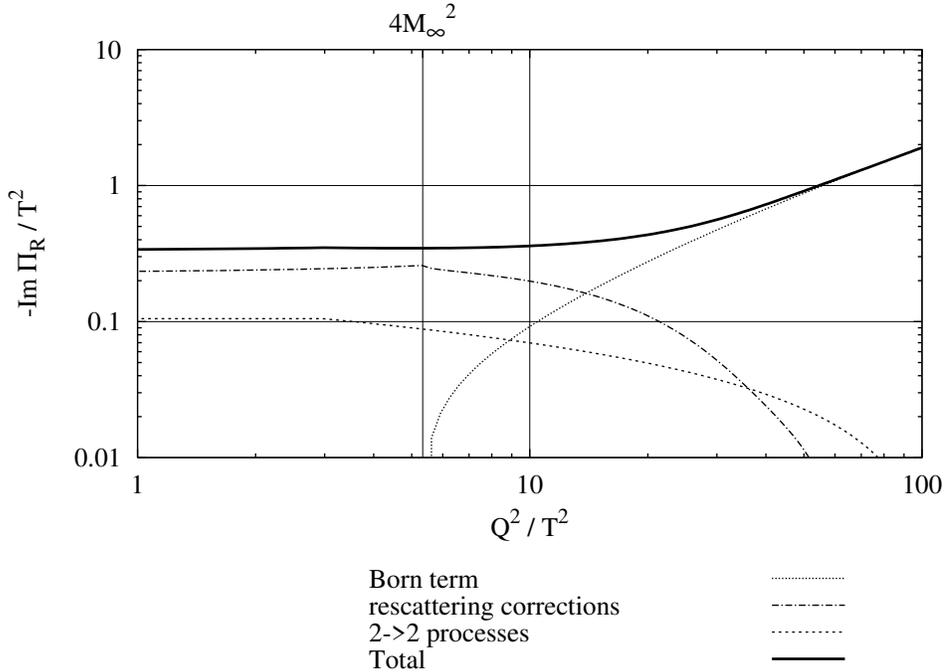}}}}
\caption{\label{fig:all} All the contributions to ${\rm Im}\,\PiR$
up to ${\cal O}(\alpha_{\rm s})$. In this plot $\alpha_{\rm s}=0.3$ and $q_0/T=50$.}
\end{figure}

\begin{figure}[htbp]
\centerline{
\resizebox*{!}{8cm}{\rotatebox{-90}{\includegraphics{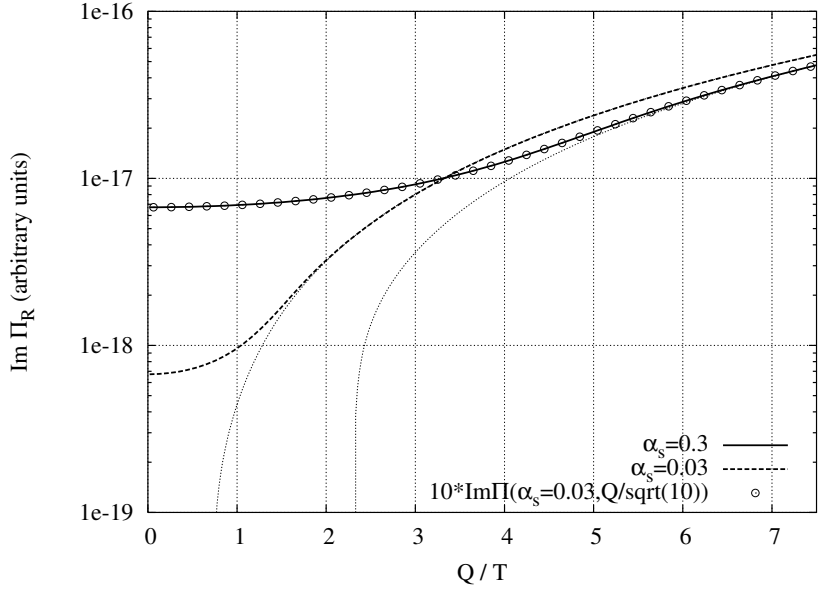}}}}
\caption{\label{fig:coupling-dep} ${\rm Im}\,\Pi_{_{R}}$ as a function
of the photon invariant mass for two values of the strong coupling
constant. The dotted lines are the corresponding Born terms. The
circles illustrate the scaling law of Eq.~(\ref{eq:scaling}). The
value of $q_0/T$ is set to $30$ in this plot.}
\end{figure}

If one now includes also the Born term, an interesting feature of the
result is that there is no trace of the threshold at
$Q^2=4M_\infty^2$. In other words, the step function
$\theta(Q^2-4M_\infty^2)$ behavior which is characteristic of the
Drell-Yan process $q\bar{q}\to \gamma^*\to l^+ l^-$ is completely washed
out when one corrects this process by in-medium rescatterings. This is
illustrated in figure \ref{fig:all}.
This property is easy to understand after the integral equations
(\ref{eq:integ-f}) and (\ref{eq:integ-g}), whose solution contains the
Born term and all the rescattering corrections, have been rewritten as
differential equations with mixed boundary conditions. The parameter
$Q^2$ enters only as a coefficient in the linear differential
equations (\ref{eq:ODE-1}).  Because of the term containing $D$, the
solution is always regular and exponentially falling at large $b$ (see
Eq.~(\ref{eq:lambda})).  This ensures that it evolves smoothly as
$Q^2$ changes\footnote{An equivalent explanation, in momentum space,
  for the smoothness at threshold is the following: in
  eqs.~(\ref{eq:integ-f}) and (\ref{eq:integ-g}) respectively, the
  $\f(\p_\perp)$ and $g(\p_\perp)$ terms under the integral originate
  from the damping rate of the fermion (due to the rescattering of the
  fermion in the medium) and can be seen to provide an imaginary part
  to $\delta E$. The pole occuring at $Q^2 > 4 \Meff^2$ in the Born
  term is then shifted away from the real axis by an amount
  proportional to the damping rate and no $\theta(-\Meff^2)$
  appears.}.  One can also add that the smoothing of the threshold
depends on the value of the strong coupling constant $\gs$, as
illustrated in figure \ref{fig:coupling-dep}.
On this plot, one can see a remnant of the threshold at small
coupling, while it completely disappears at strong coupling. We also
illustrate on this plot a simple scaling property of the function
${\rm Im}\,\Pi_{_{R}}{}^\mu_\mu(\alpha_{\rm s},q_0,Q^2,T)$; from
equations (\ref{eq:LPM-complete}),
(\ref{eq:integ-f}) and (\ref{eq:integ-g}) one can show that the
dependence on $\alphas$ scales according to
\begin{equation}
{\rm Im}\,\Pi_{_{R}}{}^\mu_\mu(\alpha_{\rm s},q_0,Q^2,T)=\alpha_{\rm s} T^2
F\left(\frac{q_0}{T},\frac{Q^2}{\alpha_{\rm s}T^2}\right)\; ,
\label{eq:scaling}
\end{equation}
where $F$ is a dimensionless function of $q_0/T$ and $Q^2/\alpha_{\rm
s}T^2$ (in other words, $gT$ is the natural unit for the photon
invariant mass in this problem).

Concerning the respective size of the transverse and longitudinal photon
polarizations, we have checked that at low $Q^2/q_0^2$, the rate is, as
expected, dominated by the transverse mode and that the longitudinal mode
becomes important only when $Q^2/q_0^2$ approaches 1, a region where
rescattering effects are not very large.

\begin{figure}[htbp]
\centerline{
\resizebox*{!}{9cm}{\rotatebox{-90}{\includegraphics{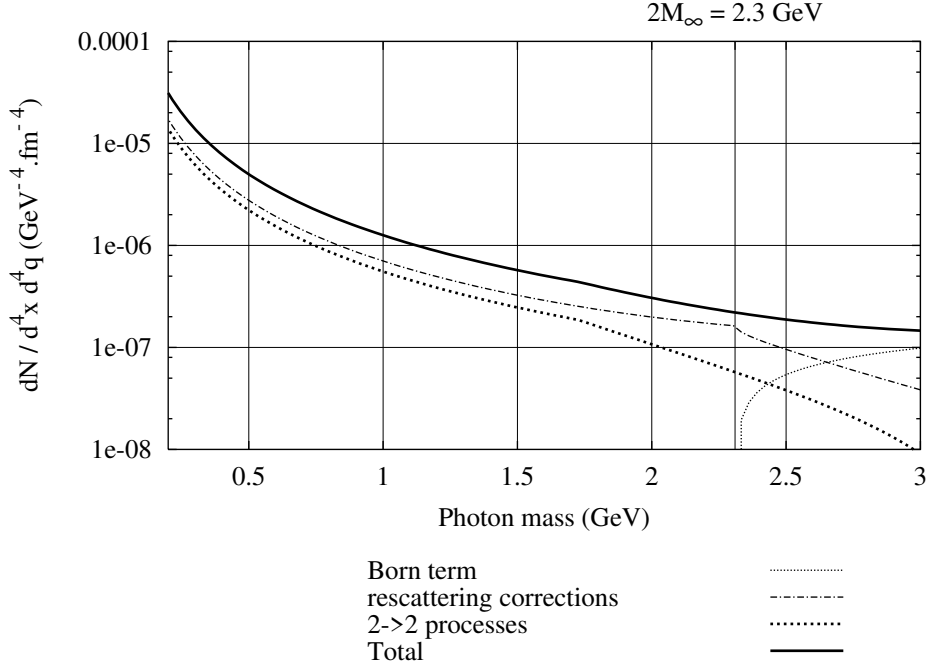}}}}
\caption{\label{fig:rate-mass} Dependence of the dilepton rate on the
mass of the pair. In this plot $T=1$~GeV, $q_0=5$~GeV, $N_{_{F}}=2$ and
$\alpha_{\rm s}=0.3$.}
\end{figure}

In order to illustrate how the LPM corrections evaluated in the present
paper affect the dilepton production rate, we present now a series of plots
for realistic values of the parameters. Figure \ref{fig:rate-mass} shows
the dependence of the dilepton rate with respect to the mass of the
lepton pair, the pair total energy being set to $5$~GeV, for a
temperature of $1$~GeV (we consider $2$ quark flavors, and take
$\alpha_{\rm s}=0.3$).
One can see on this plot that considering only the Born term and the
$2\to 2$ processes is not a good approximation. In the low mass region
($Q\le 1$~GeV), the rescattering diagrams more than double the
contribution of the $2\to 2$ processes. In the intermediate mass region,
near the tree-level threshold, the effect of these processes is even
larger and they completely wash out the threshold, as explained
before. It is only for large pair masses ($Q\ge 3$~GeV) that these
corrections start being small compared to the Born term. However, the
ratio $Q/q_0$ of the mass over the energy of the pair becomes large and the
approximations used cease to be reliable.

\begin{figure}[htbp]
\centerline{
\resizebox*{!}{8cm}{\rotatebox{-90}{\includegraphics{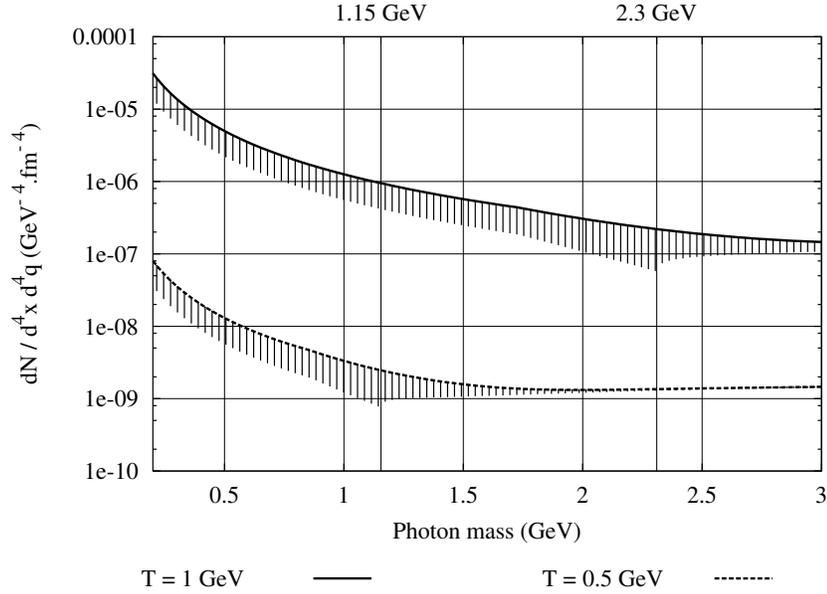}}}}
\caption{\label{fig:rate-T} Dependence of the dilepton rate on the
temperature. The vertical lines represent the contribution of the
processes involving scatterings in the medium. In this plot $q_0=5$~GeV,
$N_{_{F}}=2$ and $\alpha_{\rm s}=0.3$.}
\end{figure}

One can also study how this mass dependence varies with the
temperature. In figure \ref{fig:rate-T}, we display the total dilepton
rate for two different temperatures ($T=1$~GeV and $T=500$~MeV), the
other parameters being the same as in the previous figure.
In this plot, the vertical lines indicate how much of the total rate is
due to the rescattering contributions. In addition to the drop one
expects when one lowers the temperature, it is possible to see that the
effect of these processes are limited to smaller values of the mass
$Q$. This is because the scale which controls these effects
is controlled by thermal masses proportional to $T$.

\begin{figure}[htbp]
\centerline{
\resizebox*{!}{8cm}{\rotatebox{-90}{\includegraphics{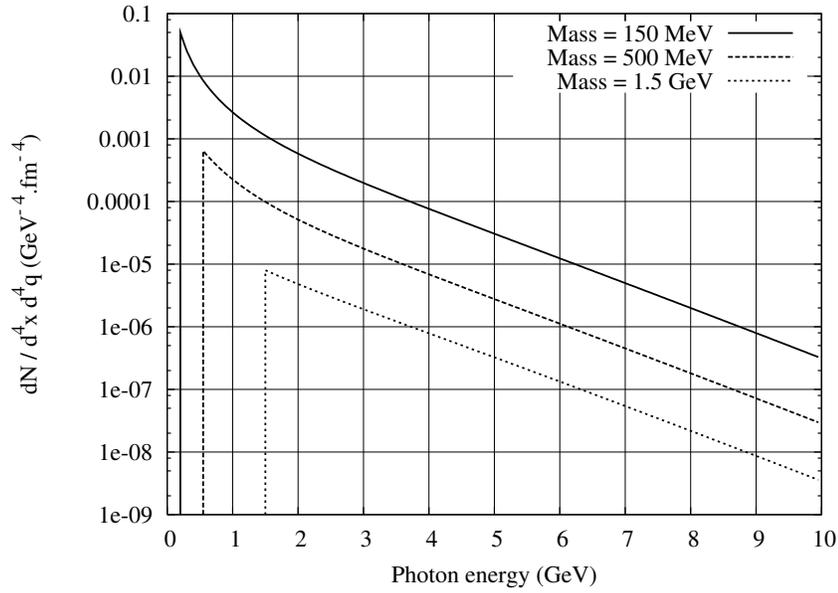}}}}
\caption{\label{fig:rate-E} Dilepton rate as a function of the pair
energy, for $Q=150$~MeV, $Q=500$~MeV and $Q=1.5$~GeV. In this plot
$T=1$~GeV, $N_{_{F}}=2$ and $\alpha_{\rm s}=0.3$.}
\end{figure}

Finally, in figure \ref{fig:rate-E}, we display the dilepton rate as a
function of the pair energy, the mass $Q$ being fixed.
At large energy, one can see the typical exponential drop in
$\exp(-q_0/T)$. More precisely, the asymptotic behavior of the
dilepton rate at large $q_0$ and fixed $Q$ is in
$\sqrt{q_0}\exp(-q_0/T)$, the factor $\sqrt{q_0}$ being characteristic
of the LPM effect at large photon energies. In addition, increasing
the pair mass has the expected effect of making the rate
decrease. Note that the curves in this plot are unreliable at the low
end, near $q_0\approx Q$, since the approximation $q_0\approx q$ is
certainly invalid in this region. This is the region where the lepton
pair is produced almost at rest in the plasma frame ($q\approx 0$),
which precludes any use of a collinear kinematics.

\section{Conclusions}

In this paper we completed the ${\cal O}(\alpha_{\rm s})$ calculation
of the rate of low-mass lepton pairs produced at large momentum in a
plasma in equilibrium. Production of the pairs in multiple scattering
processes are found to be important and even dominate the rate in a
wide range of masses. An interesting result is the absence of a
threshold in the mass dependence of the virtual photon: the
rescattering processes completely wash out the $q \bar q$ annihilation
threshold naively expected.  
The absence of a threshold in the photon
invariant mass looks different from the recent results of lattice
calculations \cite{KarscLPSW1,KarscDLPS1} where a pronounced threshold is
observed in the production of a lepton pair at rest in quark-gluon
plasma in equilibrium (in the quenched approximation). 
We emphasize, however, that we cannot directly compare our results to
these lattice results, since we consider only low invariant mass
dileptons at high 
energy, that is, $Q^\mu$ close to the light cone (in the plasma frame),
while the lattice data presented in \cite{KarscLPSW1,KarscDLPS1} are for
dileptons at rest--a regime where our treatment is not valid.

The present results assume that the plasma is in equilibrium, and
use asymptotic values for the various thermal and screening masses. They are
therefore expected to be quantitatively reliable only at extremely high
temperature. More realistic results for comparison with present and future
experiments at RHIC and LHC could be obtained by using more realistic mass
estimates, from the lattice for example, and also by relaxing the assumption of
chemical equilibrium. Of course, the obtained results would be on a less firm
theoretical ground but they might be phenomenologically useful.

The dilepton channel can offer complementary information to the real photon
channel as a probe for formation of a quark-gluon plasma in heavy ion
collisions, since the background is expected to be relatively much less
important than for real photon. In this respect it would be useful to evaluate
the contribution of non thermal sources of lepton pairs for comparison with the
rates calculated here.

\section*{Acknowledgements}
We wish to thank Peter Arnold, Dominique Schiff and Larry
Yaffe for useful conversations. F.G. would like to thank the
LPT/Orsay where part of this work has been performed.  The
work of G.M. was supported by the U.~S.~ Department of
Energy under Grant No.~DE-FG03-96ER40956.

\bibliographystyle{unsrt}

\end{document}